# Encoding Hypergraphs into Quantum States

Ri Qu, Juan Wang, Zong-shang Li, Yan-ru Bao

School of Computer Science & Technology, Tianjin University, Tianjin, China, 300072

Ionicioiu and Spiller [Phys. Rev. A 85, 062313 (2012)] have recently presented an axiomatic framework for mapping graphs to quantum states of a suitable physical system. Based on their study, we first extend the axiomatic framework to hypergraphs by means of modifying its axioms and consistency conditions. Then we use the axiomatic approach to encode hypergraphs into a new family of quantum states, called the hypergraph states. Moreover, we also try to do the followings: (i) to show that real equally weighted states, which occur in Grover and Deutsch-Joza algorithms, are equivalent to hypergraph states; (ii) to describe the relations among hypergraph states, graph states and stabilizer states; (iii) to provide some transformation rules, stated in purely hypergraph theoretical terms, which completely characterize the evolution of hypergraph states under some local operations, including operators in Pauli group and some special local Pauli measurements; and (iv) to investigate some properties of multipartite entanglement of hypergraph states by hypergraph theory.

1. Introduction

It is well known that any *graph state* [1-4] can be constructed on the basis of a (simple and undirected) graph. Although graph states can describe a large family of entangled states including *cluster states* [5], *GHZ states*, *stabilizer states* [6], etc., it is clear that they cannot represent all entangled states. To go beyond graph states and still keep the appealing connection to graphs, Ref. [7] introduces an axiomatic framework for mapping graphs to quantum states of a suitable physical system, and extends this framework to directed graphs and weighted graphs. Several classes of multipartite entangled states, such as *qudit graph states* [8], *Gaussian cluster states* [9], *projected entangled pair states* [10], and *quantum random networks* [11], emerge from the axiomatic framework. The main aim of this paper is to develop a new approach of encoding hypergraphs into quantum states. Therefore we generalize the above axiomatic framework to hypergraphs. We use the axiomatic approach to define a new class of quantum states of $n$ qubits, called the hypergraph states. Any hypergraph state can be constructed by a (undirected) hypergraph as follows. Each vertex of the hypergraph labels a qubit, and each hyperedge is associated with an operator called a hyperedge gate. The hypergraph state is constructed from some initial state by successively applying the hyperedge gate corresponding to each hyperedge of the hypergraph.

There have been several approaches of mapping (classical) Boolean functions to quantum states which include hypergraph states. Ref. [12] introduces the concept of a *quantum Boolean function* on $n$ qubits, which is defined as a unitary and hermitian operator on $n$ qubits. Clearly, each classical Boolean function $f:\{0,1\}^n \to \{0,1\}$ may be implemented on a quantum computer by means of the so-called *phase oracle* $O_f$ as follows.

$$O_f |+\rangle^{\otimes n} = |\psi_f\rangle \equiv \frac{1}{\sqrt{2^n}} \sum_{x=0}^{2^n-1} (-1)^{f(x)} |x\rangle \qquad (1)$$

where $|x\rangle$ represent the computational basis states of *n* qubits. Since the phase oracle $O_f$ is unitary and hermitian, it is a quantum Boolean function corresponding to *f*. Note that the state $|\psi_f\rangle$, called a *real equally weighted state* in [13, 14], occurs in Grover [15] and Deutsch-Joza [16] algorithms. In this paper, we show that real equally weighted states are equivalent to hypergraph states. Thus any real equally weighted state can be constructed by not only a Boolean function but also a hypergraph. It implies that we may investigate the properties of the real equally weighted states by means of the hypergraph theory. We also show that hyperedge gates are a class of special quantum Boolean functions which are independent and commutative with each other. Thus each hypergraph can give rise to a quantum Boolean function in a natural way.

Ref. [17] shows every Boolean function $f : \{0,1\}^n \to \{0,1\}$ can give rise to a so-called *quantum Boolean state* which is expressed (up to a global factor) in the form $\sum_{x \in \{0,1\}^n} |x\rangle |f(x)\rangle$. Moreover, each Boolean state can be translated into a *categorical (Boolean) tensor network state*. Note that the state $\sum_{x \in \{0,1\}^n} \langle -|f(x)\rangle |x\rangle$, where $|-\rangle \equiv \frac{1}{\sqrt{2}}(|0\rangle + |1\rangle)$, is just (up to a global scalar factor) a real equally weighted state, i.e., a hypergraph state. Boolean tensor network states are used to study the decidability questions in [18], and to define a class of algorithms for efficiently solving some search problems in [19]. Ref. [20] uses symmetries of Boolean functions to unify and extend various constructions of spin Hamiltonians embedding Boolean functions into their ground-state subspaces.

One may ask (i) whether every hypergraph state is of stabilizer states and (ii) whether every hypergraph state is of graph states. If local unitary transformations are not considered, our answers about the above questions are both "no". This implies that the relations among hypergraph states, graph states and stabilizer states are described as follows: hypergraph states include graph states; graph states constitute a subclass of the stabilizer states; and any hypergraph state, constructed by a hypergraph whose rank is more than *2*, is not of stabilizer states.

It is clear that one can characterize hypergraph states by means of hypergraph theory. In this paper we show how to translate the action of some local operations (including operators in Pauli group and some special local Pauli measurements) on hypergraph states into the transformations on their corresponding hypergraphs, that is, to derive some transformation rules, stated in purely hypergraph theoretical terms, which completely characterize the evolution of hypergraph states under these local operations. Moreover, we also study some properties of *multipartite entanglement* [21] of hypergraph states by hypergraph theory.

This paper is organized as follows. In Sec. 2 we will recall some notations of hypergraphs, Boolean functions and Pauli group. We will show how to encode hypergraphs into quantum states by an axiomatic framework in Sec. 3. In Sec. 4 we will define the hypergraph states by means of the framework in Sec.3. In Sec. 5 we will show that real equally weighted states are equivalent to hypergraph states. We will show how any operator in Pauli group on a hypergraph state is translated into the operation of adding some specified hyperedges to its corresponding hypergraph in Sec. 6. In Sec. 7 we will describe the relations among hypergraph states, graph states and stabilizer states. We will show how some special local Pauli measurements are translated into the operations of deleting the vertices in Sec. 8. We will investigate some properties of multipartite entanglement of hypergraph states in Sec. 9. We will summarize our conclusions in Sec. 10.

2. Preliminaries

Formally, a *hypergraph* is a pair $(V, E)$, where $V$ is the set of *vertices*, $E \subseteq \wp(V)$ is the set of *hyperedges* and $\wp(S)$ denotes the power set of the set $S$. The set of all hypergraphs of $n$ vertices is denoted by $\Theta_n$. The *empty hypergraph* is defined as $(V, \Phi)$. If a hypergraph only contains the *empty hyperedge* $\Phi$ or one-vertex hyperedges (called *loops*), it is *trivial*. The *rank* of a hypergraph $g$, denoted by $ran(g)$, is the maximum cardinality of a hyperedge in $g$. Moreover, a hypergraph can be depicted by the visual form as shown in Fig. 1. Each vertex is represented as a dot while each hyperedge is represented as a closed curve which encloses the dots corresponding to vertices incident with the hyperedge.

A hypergraph $(V', E')$ is called a *subhypergraph* of $(V, E)$ if $V' \subseteq V$ and $E' \subseteq E$. Two hypergraphs $(V, E)$ and $(V', E')$ are *isomorphic* if there exists a permutation $P$ on $V$, that is, a *1-1* mapping $P : V \to V'$, such that $V' = P(V)$ and $E' = \bigcup_{e \in E} \{P(v) | v \in e\}$. The vertices incident with the same hyperedge are referred to as being *adjacent*. A sequence of vertices $v_1, v_2, ..., v_p$ such that $v_k$ and $v_{k+1}$ are adjacent for all $1 \leq k \leq p-1$ is called a *path* joining $v_1$ to $v_p$. A hypergraph is *connected* if any two vertices are joined by a path. Otherwise, it is *disconnected*. A *component* of a hypergraph $g$ is a connected subhypergraph contained in no other connected subhypergraph. The number of components of a hypergraph $g$ is denoted by $con(g)$.

We define the sum of $g = (V, E)$ and $g' = (V', E')$ as $g \Delta g' \equiv (V \cup V', E \Delta E')$ where $E \Delta E'$ denotes the symmetric difference of $E$ and $E'$, that is, $E \Delta E' = E \cup E' - E \cap E'$. Given a hypergraph $g = (V, E)$, there are two ways of deleting vertex $k$ to obtain a new hypergraph $g' = (V', E')$: (i) Delete vertex $k$ in $V$, together with all hyperedges incident with $k$, which means that $V' = V - \{k\}$ and $E' = E \Delta \{e \mid k \in e \wedge e \in E\}$. Denote the obtained hypergraph $g'$ by $g -_+ \{k\}$; (ii) At first delete vertex $k$ in $V$ and replace each hyperedge $e$ incident with $k$ to $e - \{k\}$. Then delete all repeated hyperedges. This implies $V' = V - \{k\}$ and $E' = E \Delta \{e \mid k \in e \wedge e \in E\} \Delta \{e - \{k\} \mid k \in e \wedge e \in E\}$. Denote the obtained hypergraph $g'$ by $g -_- \{k\}$. In Fig.1, the hypergraphs $g_c$ and $g_d$ are respectively corresponding to $g_a -_+ \{1\}$ and $g_a -_- \{1\}$. Similar to graph theory, one can define the vertex cover of a hypergraph $g = (V, E)$. If a trivial hypergraph is obtained by deleting all vertices in a subset of $V$, then the subset is called a *vertex cover* of $g$. In Fig.1, the sets $\{3\}$ and $\{1, 4\}$ are two different vertex covers of the hypergraphs $g_a$. Moreover, for a set of the hyperedges $F \subseteq \wp(V)$, adding all hyperedges of $F$ to a hypergraph $g = (V, E)$ will obtain a new hypergraph $g + F \equiv (V, E \Delta F)$. The hypergraph $g_b = g_a + \{\{\Phi\}, \{2, 3\}\}$ is shown in Fig.1.

Let $[n] \equiv \{1, 2, ..., n\}$. Define a mapping on $\wp([n])$ as

$$\forall e \subseteq [n], c(e) = \begin{cases} 1 & e = \Phi \\ \prod_{k \in e} x_k & e \neq \Phi \end{cases}. \tag{2}$$

Then an *n*-variable Boolean function $f(x) \equiv f(x_1, x_2, ..., x_n)$ can be written as the so-called *algebraic normal form* as follows.

$$\bigoplus_{e \subseteq [n]} a_e c(e) \tag{3}$$

where the coefficients $a_e \in \{0, 1\}$ and $\oplus$ denotes the addition operator over $\mathbb{Z}_2$. The *algebraic degree* and *codegree* of $f$ are respectively defined as $\deg(f) \equiv \max_{e \in [n] \wedge a_e \neq 0} |e|$ and $\operatorname{codeg}(f) \equiv \min_{e \in [n] \wedge a_e \neq 0} |e|$ where $|e|$ denotes the cardinality of the set $e$. The *quadratic functions* are the Boolean functions with algebraic degree at most two. The set of all *n*-variable *quadratic functions* is denoted by $Q_n$. The set of all *n*-variable Boolean functions is denoted by $\Omega_n$.

For convenience, let $[n]$ be the set of $n$ vertices. Then we can construct a *1-1* mapping $u : \Theta_n \to \Omega_n$ which satisfies $\forall g = ([n], E) \in \Theta_n$,

$$u(g) = \bigoplus_{e \in E} c(e). \tag{4}$$

Moreover, it is clear that $\forall g, g' \in \Theta_n$, $u(g \Delta g') = u(g) \oplus u(g')$, which implies that $(\Theta_n, \Delta)$ is isomorphic with $(\Omega_n, \oplus)$. Thus every $n$-variable Boolean function can be uniquely represented as a hypergraph of $n$ vertices.

Denote the 2 by 2 identity matrix by $I$ and let

$$\sigma_x \equiv \begin{bmatrix} 0 & 1 \\ 1 & 0 \end{bmatrix}, \ \sigma_y \equiv \begin{bmatrix} 0 & -i \\ i & 0 \end{bmatrix}, \ \sigma_z \equiv \begin{bmatrix} 1 & 0 \\ 0 & -1 \end{bmatrix} \tag{5}$$

Define $G_n$, the *Pauli group* on $n$ qubits, as $2^n$ by $2^n$ matrices of the form $\alpha p_1 \otimes p_2 \otimes ... \otimes p_n$ for some $\alpha \in \{1, -1, i, -i\}$ and $p_k \in \{I, \sigma_x, \sigma_y, \sigma_z\}$. Define $\sigma_x^{(k)}$ as $\sigma_x$ acting on the *k-th* qubit, i.e., $I^{\otimes k-1} \otimes \sigma_x \otimes I^{\otimes n-k}$. Similarly for $\sigma_y^{(k)}$ and $\sigma_z^{(k)}$.

3. Axiomatic framework and Hypergraphs

Ref. [7] has presented an axiomatic framework for encoding graphs into quantum states. The framework defines a class of theories characterized by a triplet $(H, |\phi\rangle, U)$ where $H$ is the Hilbert space associated to a vertex, $|\phi\rangle \in H$ is an initial state, and $U$ is an edge operator acting on $H \otimes H$. A graph $g$ is mapped to the corresponding state $|g\rangle$ constructed from the initial state $|\phi\rangle^{\otimes n}$ by successively applying the edge operator corresponding to each edge in $g$.

Let $|+\rangle \equiv \frac{1}{\sqrt{2}}(|0\rangle + |1\rangle)$ and $Z_2 \equiv diag(1,1,1,-1)$. Then $(\mathbb{C}^2, |+\rangle, Z_2)$ is the simplest case and it is used to encode simple graphs into graph states. The framework has been extended to multiple graphs, random graphs, directed graphs and weighted graphs [7]. In this section, we extend the framework to hypergraphs. We firstly modify the axioms A1-A3 in [7] into the following axioms A1'- A3' in order to fit to hypergraphs.

*Axiom A1': Separability.* Suppose two hypergraphs $g = (V, E)$, $g' = (V', E')$ have no same vertex, i.e., $V \cap V' = \Phi$. Then we have

$$|g \Delta g'\rangle = |g\rangle \otimes |g'\rangle. \tag{6}$$

*Axiom A2': Hypergraph isomorphism.* If two hypergraphs $g = (V, E)$, $g' = (V', E')$ are isomorphic, the corresponding density operators $\rho = |g\rangle\langle g|$ and $\rho' = |g'\rangle\langle g'|$ satisfy

$$\rho' = D(P)\rho D(P)^{-1} \qquad (7)$$

where $D(P)$ is a matrix representation of the permutation $P$ on $V$ mapping $g$ to $g'$.

According to the above two axioms, we can obtain the following proposition.

*Proposition 1.* Given a hypergraph $g=(V,E)$ of $n$ vertices, the corresponding quantum state $|g\rangle$ belongs to a Hilbert space $H^{\otimes n}$ of $n$ identical quantum systems where $H$ is the Hilbert space associated to a single vertex. Moreover, the empty hypergraph is mapped to $|\phi\rangle^{\otimes n}$ where $|\phi\rangle \in H$.

*Proof.* It is the same as proposition 1 in [7]. □

*Axiom A3': Universal hyperedge operator.* If two hypergraphs $g=(V,E)$, $g'=(V,E')$ differ by a single hyperedge, i.e., $E' = E \cup \{e\}$ where $e \subseteq V$, then

$$|g'\rangle = U_e |g\rangle. \qquad (8)$$

The hyperedge operator $U_e$ is independent of $g$ and $g'$ and depends only on the hyperedge $e$.

Given a hypergraph $g=(V,E)$ of $n$ vertices, the axioms A1'-A3' provide a constructive way to obtain the corresponding quantum state $|g\rangle$: starting from the empty hypergraph $|\phi\rangle^{\otimes n}$ we apply the hyperedge operator corresponding to each hyperedge in $E$, that is,

$$|g\rangle = \prod_{e \in E} U_e |\phi\rangle^{\otimes n}. \qquad (9)$$

This construction is consistent if the hyperedge operator satisfies the following conditions C1'-C3' which are a generalization of three consistency conditions in [7]. Since there are $n+1$ types of hyperedges according to the cardinality of a hyperedge, we give the first condition.

*Condition C1': Variety.* There are $n+1$ basic operators $U_0, U_1, ..., U_n$ where the operator $U_k$ acts on $H^{\otimes k}$ for all $1 \leq k \leq n$ and $U_0 \in \mathbb{C}$ is a constant.

In this paper, we only discuss the undirected hypergraphs. Hence we combine the conditions C2-C3 shown in [7] into the following condition.

*Condition C2': Locality and symmetry.* The hyperedge operator $U_e$ acts nontrivially only on $H^{\otimes |e|}$ associated with vertices in $e$,

$$U_e = U_{|e|} \otimes I^{\otimes n - |e|}, \qquad (10)$$

with $I$ the identify acting on the rest. In particular, $U_\Phi = U_0 \cdot I^{\otimes n}$.

*Condition C3': Commutability.* Any two hyperedge operators commute.

We denote the above framework for hypergraphs by a triple $(H, |\phi\rangle, \{U_k | 0 \le k \le n\})$. In the following section, we will identify $H$, $|\phi\rangle$ and $\{U_k | 0 \le k \le n\}$ to construct hypergraph states.

4. Hypergraph states

We firstly define some special gates called the hyperedge gates and discuss their properties. Let $Z_k$ be the $2^k$ by $2^k$ diagonal matrix which satisfies

$$(Z_k)_{jj} = \begin{cases} -1 & j = 2^k \\ 1 & others \end{cases}. \tag{11}$$

Clearly, $Z_0$ is equal to $-1$ and $Z_1$ is just the Pauli matrix $\sigma_z$. If $k \ge 2$, then $Z_k$ is a controlled phase operator acting on $k$ qubits with $k$-$1$ control qubits and one target qubit. Note that $Z_k$ does not depend on which of $k$ qubits is the target one.

Suppose $e \subseteq [n]$. If $e = \{i_1, i_2, ..., i_k\}$, define the operator $Z_e$ as $Z_k \otimes I^{\otimes n-k}$ which means that $Z_k$ acts on the $i_1$-th, $i_2$-th, ..., and $i_k$-th qubits while $I$ acts on the rest respectively. We call $Z_e$ a *hyperedge gate* on $n$ qubits. Moreover, define the operator $Z_\Phi$ as $-I^{\otimes n}$. Note that $Z_{\{j\}}$ implies $Z_1$ acting on the $j$-th qubit, i.e., $Z_{\{j\}} = \sigma_z^{(j)} \in G_n$.

Clearly, hyperedge gates have the following properties. $\forall e, e' \subseteq [n]$, (i) $Z_e$ is hermitian and unitary, thus $Z_e^2 = I^{\otimes n}$; (ii) $Z_e Z_{e'} = Z_{e'} Z_e$; (iii) $Z_e$ is independent with the other hyperedge gates, i.e., $Z_e$ can not be written by a product expression of the other hyperedge gates. Note that hyperedge gates are a class of special quantum Boolean functions which are independent and commutative with each other. This implies that any hypergraph $([n], E)$ can give rise to a quantum Boolean function $\prod_{e \in E} Z_e$.

Given a hypergraph $g = ([n], E)$, an $n$-qubit state $|g\rangle$ can be constructed by $g$ as follows. Each vertex labels a qubit (i.e., $H = \mathbb{C}^2$) initialized in $|\phi\rangle = |+\rangle$. The basic operators are $Z_k$ for all $0 \leq k \leq n$. The state $|g\rangle$ is obtained from the initial state $|+\rangle^{\otimes n}$ by applying the hyperedge operator $Z_e$ for each hyperedge $e \in E$, that is,

$$|g\rangle = \prod_{e \in E} Z_e |+\rangle^{\otimes n}. \tag{12}$$

This construction is consistent since the basic operators $\{Z_k \mid 0 \leq k \leq n\}$ and the hyperedge gates satisfy the conditions C1'-C3'. The state $|g\rangle$ is called an *n*-qubit *hypergraph state*. Thus hypergraph states are corresponding to $(\mathbb{C}^2, |+\rangle, \{Z_k \mid 0 \leq k \leq n\})$.

5. Real equally weighted states and hypergraph states

It is known that real equally weighted states, which are of the form $|\psi_f\rangle$ in (1), occur in Grover and Deutsch-Joza algorithms. In this section we discuss the relation between real equally weighted states and hypergraph states.

We can construct a *1-1* correspondence between $\{|g\rangle \mid g \in \Theta_n\}$ and $\{|\psi_f\rangle \mid f \in \Omega_n\}$. In fact, it is clear that $\forall e, e' \subseteq [n]$,

$$Z_e |+\rangle^{\otimes n} = \frac{1}{\sqrt{2^n}} \sum_{x=0}^{2^n-1} (-1)^{c(e)} |x\rangle \tag{13}$$

and

$$Z_e Z_{e'} |+\rangle^{\otimes n} = \frac{1}{\sqrt{2^n}} \sum_{x=0}^{2^n-1} (-1)^{c(e) \oplus c(e')} |x\rangle. \tag{14}$$

By (4) and (12), we have $\forall g = ([n], E) \in \Theta_n$,

$$|g\rangle = \prod_{e \in E} Z_e |+\rangle^{\otimes n} = \frac{1}{\sqrt{2^n}} \sum_{x=0}^{2^n-1} (-1)^{\bigoplus_{e \in E} c(e)} |x\rangle = |\psi_{u(g)}\rangle. \tag{15}$$

Hence this implies that real equally weighted states are equivalent to hypergraph states. In particular, the set of all graph states is just the set $\{|\psi_f\rangle \mid f \in Q_n \wedge \mathrm{codeg}(f) = 2\}$. This implies that hypergraph states include graph states.

6. Operators in Pauli group and adding hyperedges

In this section we show that any operator in Pauli group on a hypergraph state can be translated into the operation of adding some specified hyperedges to the corresponding hypergraph.

*Proposition 2.* Let $g = ([n], E)$ be a hypergraph, and let $U \in \{\sigma_x^{(k)}, \sigma_z^{(k)}\}$ where $k \in [n]$.

Then the state $|g'\rangle = U|g\rangle$ is corresponding to the hypergraph $g' = ([n], E')$ where

$$E' = \begin{cases} E\Delta\{e-\{k\} | k \in e \wedge e \in E\} & U = \sigma_x^{(k)} \\ E\Delta\{\{k\}\} & U = \sigma_z^{(k)} \end{cases}. \quad (16)$$

It implies that $\sigma_x^{(k)}$ responds to adding all hyperedges of $\{e-\{k\} | k \in e \wedge e \in E\}$ to $g$ while $\sigma_z^{(k)}$ is corresponding to adding the hyperedge $\{k\}$ to $g$. In Fig. 1, the hypergraph $g_a$ is transformed into $g_b$ by applying $\sigma_x^{(1)}$ on the hypergraph state $|g_a\rangle$.

*Proof.* From (15), we have $|g\rangle = |\psi_{u(g)}\rangle$ and $u(g)(x) = \underset{e \in E}{\oplus} c(e)$. It is well known that each $f \in \Omega_n$ can be written as

$$f(x) = (1-x_k) \cdot f(x_1, ..., x_{k-1}, 0, x_{k+1}, ..., x_n) + x_k \cdot f(x_1, ..., x_{k-1}, 1, x_{k+1}, ..., x_n). \quad (17)$$

Thus we can obtain

$$|g\rangle = \frac{1}{\sqrt{2^n}} \sum_{x_1=0}^{1} \cdots \sum_{x_{k-1}=0}^{1} \sum_{x_{k+1}=0}^{1} \cdots \sum_{x_n=0}^{1} (-1)^{u(g)(x_1,...,x_{k-1},0,x_{k+1},...,x_n)} |x_1...x_{k-1}x_{k+1}...x_n\rangle |0\rangle^{(k)}$$

$$+ \frac{1}{\sqrt{2^n}} \sum_{x_1=0}^{1} \cdots \sum_{x_{k-1}=0}^{1} \sum_{x_{k+1}=0}^{1} \cdots \sum_{x_n=0}^{1} (-1)^{u(g)(x_1,...,x_{k-1},1,x_{k+1},...,x_n)} |x_1...x_{k-1}x_{k+1}...x_n\rangle |1\rangle^{(k)}. \quad (18)$$

When $U = \sigma_x^{(k)}$, we have

$$|g'\rangle = \sigma_x^{(k)} |g\rangle$$

$$= \frac{1}{\sqrt{2^n}} \sum_{x_1=0}^{1} \cdots \sum_{x_{k-1}=0}^{1} \sum_{x_{k+1}=0}^{1} \cdots \sum_{x_n=0}^{1} (-1)^{u(g)(x_1,...,x_{k-1},0,x_{k+1},...,x_n)} |x_1...x_{k-1}x_{k+1}...x_n\rangle |1\rangle^{(k)}$$

$$+ \frac{1}{\sqrt{2^n}} \sum_{x_1=0}^{1} \cdots \sum_{x_{k-1}=0}^{1} \sum_{x_{k+1}=0}^{1} \cdots \sum_{x_n=0}^{1} (-1)^{u(g)(x_1,...,x_{k-1},1,x_{k+1},...,x_n)} |x_1...x_{k-1}x_{k+1}...x_n\rangle |0\rangle^{(k)}$$

$$= \frac{1}{\sqrt{2^n}} \sum_{x_1=0}^{1} \sum_{x_2=0}^{1} \cdots \sum_{x_n=0}^{1} (-1)^{u(g)(x_1,...,x_{k-1},1 \oplus x_k,x_{k+1},...,x_n)} |x_1 x_2...x_n\rangle. \quad (19)$$

It is clear that $u(g)(x_1, ..., x_{k-1}, 1 \oplus x_k, x_{k+1}, ..., x_n) = \underset{e' \in E\Delta\{e-\{k\}|k \in e \wedge e \in E\}}{\oplus} c(e')$. Thus we can get $E' = E\Delta\{e-\{k\} | k \in e \wedge e \in E\}$. When $U = \sigma_z^{(k)}$, we have $E' = E\Delta\{\{k\}\}$ in that $\sigma_z^{(k)} = Z_{\{k\}}$. □

It is well known that the Pauli group $G_n$ can be generated by the set

$\{iI^{\otimes n}\} \cup \{\sigma_x^{(k)}, \sigma_z^{(k)} | k \in [n]\}$. This means that any operator $U \in G_n$ transforms a hypergraph into a new one (up to the global phase $i$) according to the above proposition. Moreover, $U$ is translated into the operation of adding the associated hyperedges.

7. Hypergraph states, graph states and stabilizer states

In this section we answer the following questions: (i) whether every hypergraph state is of stabilizer states and (ii) whether every hypergraph state is of graph states. Thereby, we describe the relations among stabilizer states, graph states and hypergraph states. Note that local unitary transformations are not considered in this section.

An $n$-qubit *stabilizer state* $|\varphi\rangle$ is defined as a simultaneous eigenstate with eigenvalue *1* of $n$ commuting and independent operators in the Pauli group $G_n$. The set $\{M | M|\varphi\rangle = |\varphi\rangle \wedge M \in G_n\}$ is called the stabilizer of $|\varphi\rangle$. We first answer the question (i) as follows.

*Propositon 3.* If the rank of a hypergraph $g = ([n], E)$ is more than *2*, then the corresponding hypergraph state $|g\rangle$ is not any stabilizer state.

*Proof.* According to the definition of the hypergraph states, we can obtain $|g\rangle = \prod_{e \in E} Z_e |+\rangle^{\otimes n}$. Clearly, the state $|+\rangle^{\otimes n}$ is of stabilizer states, and its stabilizer is generated by the set $\{\sigma_x^{(k)} | k \in [n]\}$. Assume that $|g\rangle$ is a stabilizer state. It would be true that $\left(\prod_{e \in E} Z_e\right) \sigma_x^{(k)} \left(\prod_{e \in E} Z_e\right)^\dagger \in G_n$ for all $k \in [n]$. Since $ran(g) > 2$, there exist a hyperedge $e_r$ and an integer $j$ such that $|e_r| = ran(g) > 2$ and $j \in e_r$. According to proposition 2, we have

$$\left(\prod_{e \in E} Z_e\right) \sigma_x^{(j)} \left(\prod_{e \in E} Z_e\right)^\dagger |+\rangle^{\otimes n}$$

$$= \prod_{e \in E} Z_e \cdot \prod_{e \in E \Delta \{e'-\{j\} | j \in e' \wedge e' \in E\}} Z_e |+\rangle^{\otimes n}$$

$$= \prod_{e \in \{e'-\{j\} | j \in e' \wedge e' \in E\}} Z_e |+\rangle^{\otimes n}. \tag{20}$$

Clearly, $e_r - \{j\} \in \{e' - \{j\} | j \in e' \wedge e' \in E\}$ and $|e_r - \{j\}| \geq 2$. Thus $\prod_{e \in \{e'-\{j\} | j \in e' \wedge e' \in E\}} Z_e \notin G_n$

which is contradictory with $\left(\prod_{e\in E} Z_e\right)\sigma_x^{(j)}\left(\prod_{e\in E} Z_e\right)^\dagger \in G_n$. □

As all graph states constitute a subclass of stabilizer states, the above hypergraph state is also not of graph states. Thus we can describe the relations among hypergraph states, graph states and stabilizer states: hypergraph states include graph states; graph states constitute a subclass of the stabilizer states; and any hypergraph state, constructed by a hypergraph whose rank is more than *2*, is not of stabilizer states. This is shown in Fig. 2.

8. Local Pauli measurements and deleting vertices

It is well known that any projective measurement associated with operators in the Pauli group can be treated within the stabilizer formalism [6]. Since any stabilizer state is local equivalent to a graph state [22], any measurement of operators in the Pauli group turns a given graph state into a new one (up to local unitaries). It can not be directly generalized to hypergraph states because some hypergraph states are not of stabilizer states according to the above section. However, we find the Pauli measurement of $\sigma_z^{(k)}$ can turn a given hypergraph state into another one.

*Proposition 4.* Let $g = ([n], E)$ be a hypergraph, and let $|g\rangle$ be its hypergraph state. If a local measurement of $\sigma_z^{(k)}$ on the qubit associated with vertex $k \in [n]$ is performed, then the resulting state, depending on the outcome $j \in \{1, -1\}$, is given by

$$\begin{cases} |0\rangle^{(k)}\langle 0|g\rangle = |g -_+ \{k\}\rangle \otimes |0\rangle^{(k)} & j = 1 \\ |1\rangle^{(k)}\langle 1|g\rangle = |g -_- \{k\}\rangle \otimes |1\rangle^{(k)} & j = -1 \end{cases}. \quad (21)$$

It implies that two possible resulting states are respectively corresponding to two ways of deleting the vertex *k*.

*Proof.* According to (15), $|g\rangle = |\psi_{u(g)}\rangle$ and $u(g)(x) = \bigoplus_{e\in E} c(e)$. Clearly, we also obtain the expression (18). It is known that

$$u(g)(x_1,...,x_{k-1},0,x_{k+1},...,x_n) = \bigoplus_{e'\in E\Delta\{e|k\in e\wedge e\in E\}} c(e') \quad (22)$$

and

$$u(g)(x_1,...,x_{k-1},1,x_{k+1},...,x_n) = \bigoplus_{e'\in E\Delta\{e|k\in e\wedge e\in E\}\Delta\{e-\{k\}|k\in e\wedge e\in E\}} c(e'), \quad (23)$$

which combined with (18) gives (21). □

Fig. 1 shows that the hypergraph $g_a$ is transformed into one of $g_c$ (when the outcome is *1*) and $g_d$ (when the outcome is *-1*) by performing a local measurement of $\sigma_z^{(1)}$ on the state $|g_a\rangle$. Note that there exists a relation between $|g -_+ \{k\}\rangle$ and $|g -_- \{k\}\rangle$: let $|g'\rangle = \delta_x^{(k)}|g\rangle$, then

$|g-_{\{k\}}\rangle = |g'-_{+\{k\}}\rangle$. By the above proposition, it is known that any measurement of the set generated by $\{\sigma_z^{(k)} | k \in [n]\} \cup \{-I^{\otimes n}\}$ can turn a given hypergraph state into another one.

9. Entanglement

Entanglement is a major resource in quantum information processing. In this section, we first investigate the entanglement structure of hypergraph states. Then we quantify the entanglement in hypergraph states by the *Schmidt measure* [23] and give some properties.

A pure state is *m-separable* if it can be written as tensor products of pure states of *m* subsystems [21]. An *n*-qubit pure state is *fully separable* if it is *n*-separable. Otherwise, it is entangled. If an *n*-qubit pure state is not biseparable, it is *completely entangled*. The entanglement structure of real equally weighted states has been investigated in [13] and [14]. In this section, we study the entanglement structure of hypergraph states by hypergraph theoretical terms.

*Proposition 5.* Let $g = ([n], E)$ be a hypergraph. Then the state $|g\rangle$ is *m*-separable if and only if there is *m* subhypergraphs $g_1 = (V_1, E_1), g_2 = (V_2, E_2), ..., g_m = (V_m, E_m)$ such that $\{V_1, V_2, ...V_m\}$ is a partition of $[n]$, $E = \bigcup_{1 \leq k \leq m} E_k$ and $|g\rangle = \bigotimes_{1 \leq k \leq m} |g_k\rangle$.

*Proof.* (i) "if". It is clear that $E_j \cap E_k = \Phi$ for all $j \neq k$. Thus we have

$$|g\rangle = \prod_{e \in E} Z_e |+\rangle^{\otimes n} = \prod_{e \in E_1} Z_e |+\rangle^{\otimes |V_1|} \otimes \prod_{e \in E_2} Z_e |+\rangle^{\otimes |V_2|} \otimes ... \otimes \prod_{e \in E_m} Z_e |+\rangle^{\otimes |V_m|}. \quad (24)$$

(ii) "only if". Since $|g\rangle$ is *m*-separable, the state $|g\rangle$ can be written as $|\varphi_1\rangle \otimes |\varphi_2\rangle \otimes ... \otimes |\varphi_m\rangle$ where $|\varphi_1\rangle, |\varphi_2\rangle, ..., |\varphi_m\rangle$ are the real equally weighted states [14]. The states $|\varphi_1\rangle, |\varphi_2\rangle, ..., |\varphi_m\rangle$ are also of hypergraph states according to Sec. 5. □

From the above proposition, one can glimpse the entanglement structure of any hypergraph state by merely looking at its corresponding hypergraph. This is shown in the following corollary.

*Corollary 6.* Let $g = ([n], E)$ be a hypergraph. Then (i) The state $|g\rangle$ is fully separable if and only if the hypergraph *g* is trivial; (ii) If the state $|g\rangle$ is completely entangled if and only if the hypergraph *g* is connected; (iii) If the state $|g\rangle$ is *m*-separable, then $m \leq con(g)$; (iv) If the state $|g\rangle$ is *m*-separable, then $m \leq \min\{n - ran(g) + 1, n\}$; (v) If *E* contains the hyperedge $[n]$, then the state $|g\rangle$ is completely entangled.

From (v) in the above corollary, the number of completely entangled states in $\{|g\rangle | g \in \Theta_n\}$

is at least $2^{2^n-1}$ since the number of hypergraphs containing the hyperedge $[n]$ is $2^{2^n-1}$. Moreover, when $n \to \infty$, the states in $\{|g\rangle \mid g \in \Theta_n\}$ are nearly completely entangled, that is, the hypergraphs of $\Theta_n$ are almost connected. In fact, the number of disconnected hypergraphs of $\Theta_n$, denoted by $N_{discon}$, satisfies

$$N_{discon} \leq \sum_{k=1}^{n-1} B(n,k) 2^{2^k} \cdot 2^{2^{n-k}} \leq \sum_{k=1}^{n-1} B(n,k) 2^{2^{n-1}+2} = (2^n - 2) 2^{2^{n-1}+2} \quad (25)$$

where $B$ denotes the binomial coefficient. It is known that $|\Theta_n| = 2^{2^n}$. Thus we have

$$\lim_{n \to \infty} \frac{N_{discon}}{2^{2^n}} \leq \lim_{n \to \infty} \frac{(2^n - 2) 2^{2^{n-1}+2}}{2^{2^n}} = 0. \quad (26)$$

It means that for large $n$ some quantum algorithms (e.g., the Grover search algorithm) typically employ complete entanglement.

Ref. [1] has characterized and quantified the multipartite entanglement of graph states in terms of the Schmidt measure. The Schmidt measure $E_S$ has the following properties [1, 23]: (i) $E_S(|\varphi\rangle) = 0$ if and only if $|\varphi\rangle$ is a product (i.e., fully separable) state; (ii) $E_S$ is an entanglement monotone that does not increase under stochastic local operations with classical communications (SLOCC). In particular, $E_S$ is invariable under local unitaries; (iii) $E_S(|\varphi\rangle + |\phi\rangle) \leq \max\{E_S(|\varphi\rangle), E_S(|\phi\rangle)\} + 1$; (iv) $E_S(|\varphi\rangle \otimes |\phi\rangle) = E_S(|\varphi\rangle)$ if $|\phi\rangle$ is fully separable.

Let $g = ([n], E)$ be a hypergraph. Then we can obtain the following propositions according to these properties of the Schmidt measure.

*Proposition 7.* Let $F = \{e - \{k\} \mid k \in e \wedge e \in E\}$ where $k \in [n]$. Then we have $E_s(|g\rangle) = E_s[|g'\rangle]$ where $g' = ([n], E \Delta F)$.

*Proof.* It is known that $|g'\rangle = \sigma_x^{(k)}|g\rangle$ by proposition 2. Thus $E_s(|g\rangle) = E_s[|g'\rangle]$.

*Proposition 8.* Let $E_s^k = \max\{E_s(|g -_+ \{k\}\rangle), E_s(|g -_- \{k\}\rangle)\}$ where $k \in [n]$. Then we have $E_s^k \leq E_s(|g\rangle) \leq E_s^k + 1$.

*Proof.* Clearly, $E_s(|g -_+ \{k\}\rangle) \leq E_s(|g\rangle)$ and $E_s(|g -_- \{k\}\rangle) \leq E_s(|g\rangle)$ by proposition 4.

Thus $E_s^k \leq E_s(|g\rangle)$. It is known that $|g\rangle = \frac{1}{\sqrt{2}}\left(|g-_+\{k\}\rangle \otimes |0\rangle^{(k)} + |g-_-\{k\}\rangle \otimes |1\rangle^{(k)}\right)$, which implies $E_s(|g\rangle) \leq E_s^k + 1$.

According to the above proposition and the definition of vertex cover, the following proposition can be easily obtained.

*Proposition 9.* The Schmidt measure of the hypergraph state $|g\rangle$ is less than or equal to the cardinality of minimal vertex cover of the hypergraph *g*.

10. Conclusion

In this paper we develop an axiomatic framework for encoding hypergraphs into quantum states, which implies that hypergraphs will play an important role in characterizing several families of multipartite quantum states. To see this, we use the axiomatic framework to define the hypergraph states which are equivalent to real equally weighted states. Therefore one can investigate the properties of real equally weighted states by hypergraph theory. For instance, we provide some transformation rules, stated in purely hypergraph theoretical terms, which completely characterize the evolution of hypergraph states under some local operations, including operators in Pauli group and some special local Pauli measurements. Moreover, we also investigate some properties of multipartite entanglement of hypergraph states by means of hypergraph theory.

**ACKNOWLEDGMENTS**

This work was financially supported by the National Natural Science Foundation of China under Grant No. 61170178. The authors would like to thank the anonymous referee whose comments improved both the editorial and technical quality of this paper.

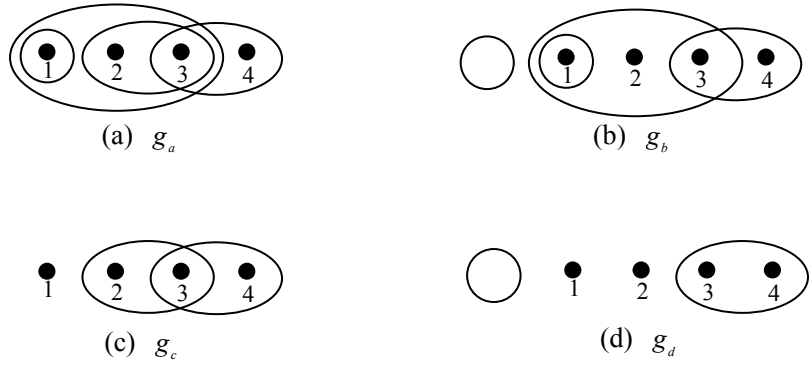

Figure 1. Examples of hypergraphs. The hypergraphs (a)-(d) have the same vertex set $\{1,2,3,4\}$. The hypergraph $g_a$ in (a) has *4* hyperedges: $\{1\}$, $\{2,3\}$, $\{3,4\}$ and $\{1,2,3\}$. In (b), the hypergraph $g_b$ also has *4* hyperedges: $\{\Phi\}$, $\{1\}$, $\{3,4\}$ and $\{1,2,3\}$. Two hyperedges, i.e., $\{2,3\}$ and $\{3,4\}$, constitute the hyperedge set of $g_c$ in (c). The hypergraph $g_d$ in (d) has 2 hyperedges: $\{\Phi\}$ and $\{3,4\}$. The hypergraphs $g_c$ and $g_d$ are respectively corresponding to $g_a -_+ \{1\}$ and $g_a -_- \{1\}$ while $g_b = g_a + \{\{\Phi\},\{2,3\}\}$.

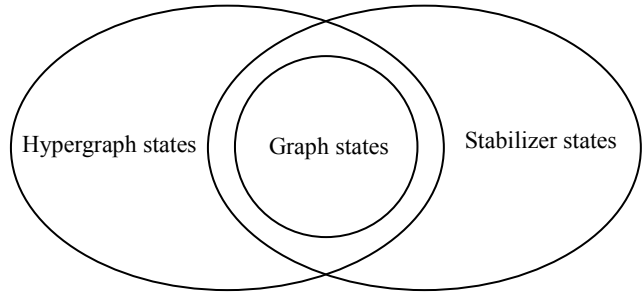

Figure 2. The relations among hypergraph states, graph states and stabilizer states. Graph states constitute a subclass of hypergraph states or stabilizer states. The intersection of hypergraph states and stabilizer states of $n$ qubits is the set $\{|g\rangle \mid g \in \Theta_n \wedge ran(g) \leq 2\}$.